\input{aipcheck}

\documentclass[
    ,final            % use final for the camera ready runs
  ]
  {aipproc}

\layoutstyle{6x9}

\begin{document}

\title{Hadron Chemistry at High $P_T$ with 
Identified Particles}


\author{Rainer J.\ Fries}{
  address={Cyclotron Institute, Texas A\&M University, College Station, TX
  77845, USA \\ and \\ RIKEN/BNL Research Center, Brookhaven National
  Laboratory, Upton NY 11973, USA}
}

\begin{abstract}
We discuss mechanisms that change the hadron chemistry for high momentum
particles emitted in high energy nuclear collisions. We argue that 
particle ratios naturally tend to be different from jets in the vacuum.
We show results of computations in a model that propagates leading
particles through a quark gluon plasma and permits elastic flavor changing
processes. We predict less suppression for kaons compared to pions in
central collision.
We also discuss elliptic flow resulting from flavor changing processes.
\end{abstract}

\maketitle

%%%%%%%%%%%%%%%%%%%%%%%%%%%%%%%%%%%%%%%%%%%%
%% MAINMATTER
%%%%%%%%%%%%%%%%%%%%%%%%%%%%%%%%%%%%%%%%%%%%

QCD jets have been used as probes of hot nuclear matter since the start
of the experimental program at the Relativistic Heavy Ion Collider (RHIC).
The first results from RHIC showed a huge suppression of single inclusive
particle yields at high transverse momenta, consistent with a large 
energy loss of fast partons in the quark gluon plasma created in the 
collisions. This quenching of jets is often parameterized by the rate of 
squared momentum transferred from the medium to a parton traversing it, 
$\hat q = \mu^2/\lambda$ where $\lambda$ is the mean free path of the 
parton in the plasma 
\cite{Wang:1991xy,BDMPS:96,Zakharov:96,Wiedemann:2000tf,gyulassy,
AMY:02}.

It was recently pointed out that the chemical composition of jets is
expected to change from the vacuum as well. Two models have been proposed
to describe this effect. In a leading particle picture one considers
an ensemble of jets described by their leading parton which propagates 
through quark gluon plasma while interacting with it. The chemical composition
of this jet or leading particle ensemble changes through flavor changing
scatterings. E.g., Compton and annihilation reactions like
 $q+g \leftrightarrow g+q$ and $q+\bar q \leftrightarrow g+g$ can change 
quark jets into gluon jets and vice versa.
Leading partons are fragmented once they are outside the medium which
translates a changing parton chemistry into a hadron chemistry which is
different from $p+p$ collisions
\cite{fries1,weiliu1,cmk,weiliu,Fries:2009tm}.

In a second approach one can study the changing chemical composition
inside a single jet cone that comes from increased multiplicities.
Additional induced radiation inside a jet cone is more favorable for the
creation of baryons and kaons compared to the vacuum \cite{Sapeta:2007ad}. 
It is 
clear that both mechanisms play a role and a complete description
of data would successfully implement a chemical coupling to the medium and
increased parton multiplicities. Here, we will focus on the former model.

\begin{figure}[t]
\centerline{
\includegraphics[width=5.5cm,angle=-90]{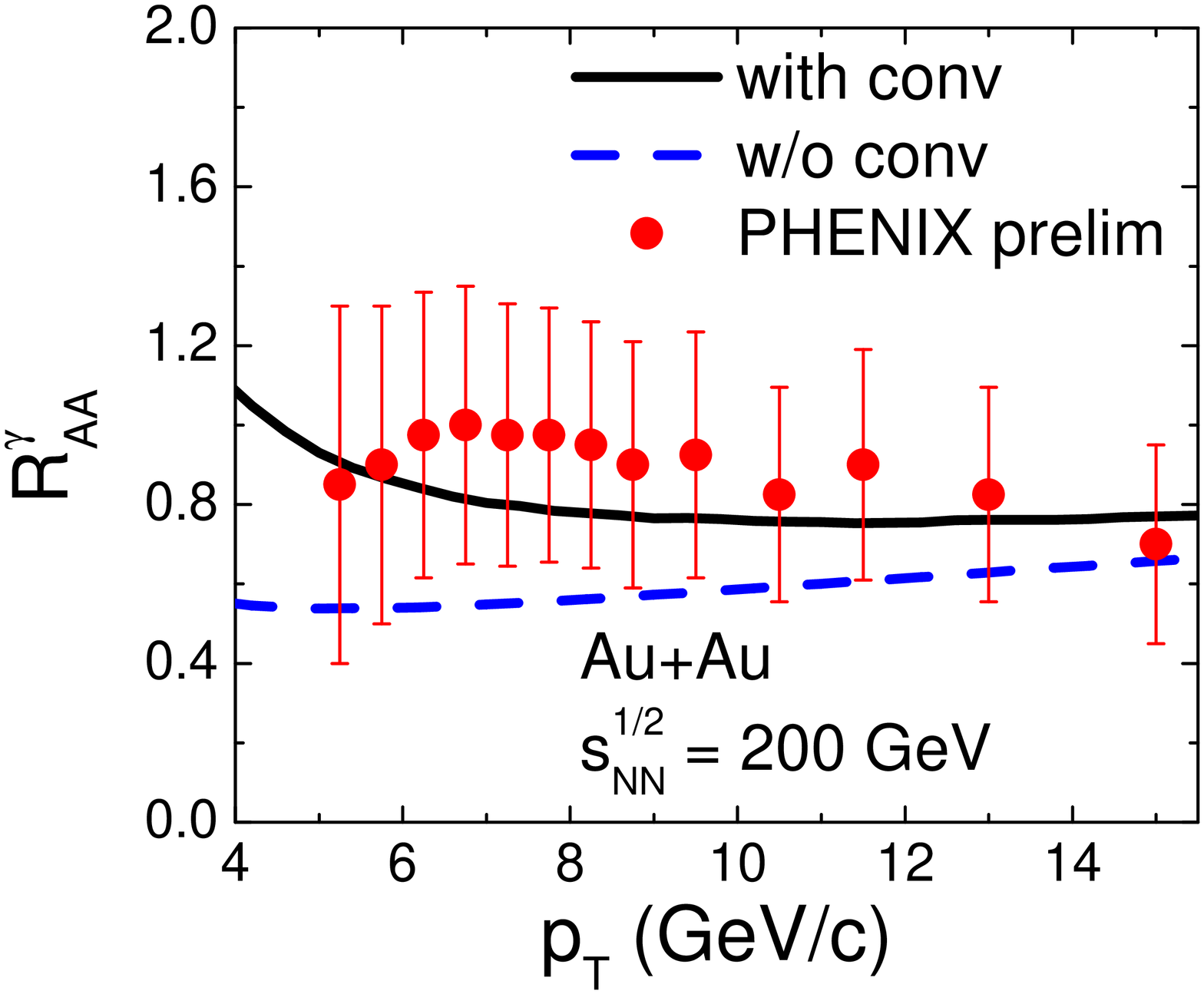}
\hspace{0.2cm}
\includegraphics[width=5.5cm,angle=-90]{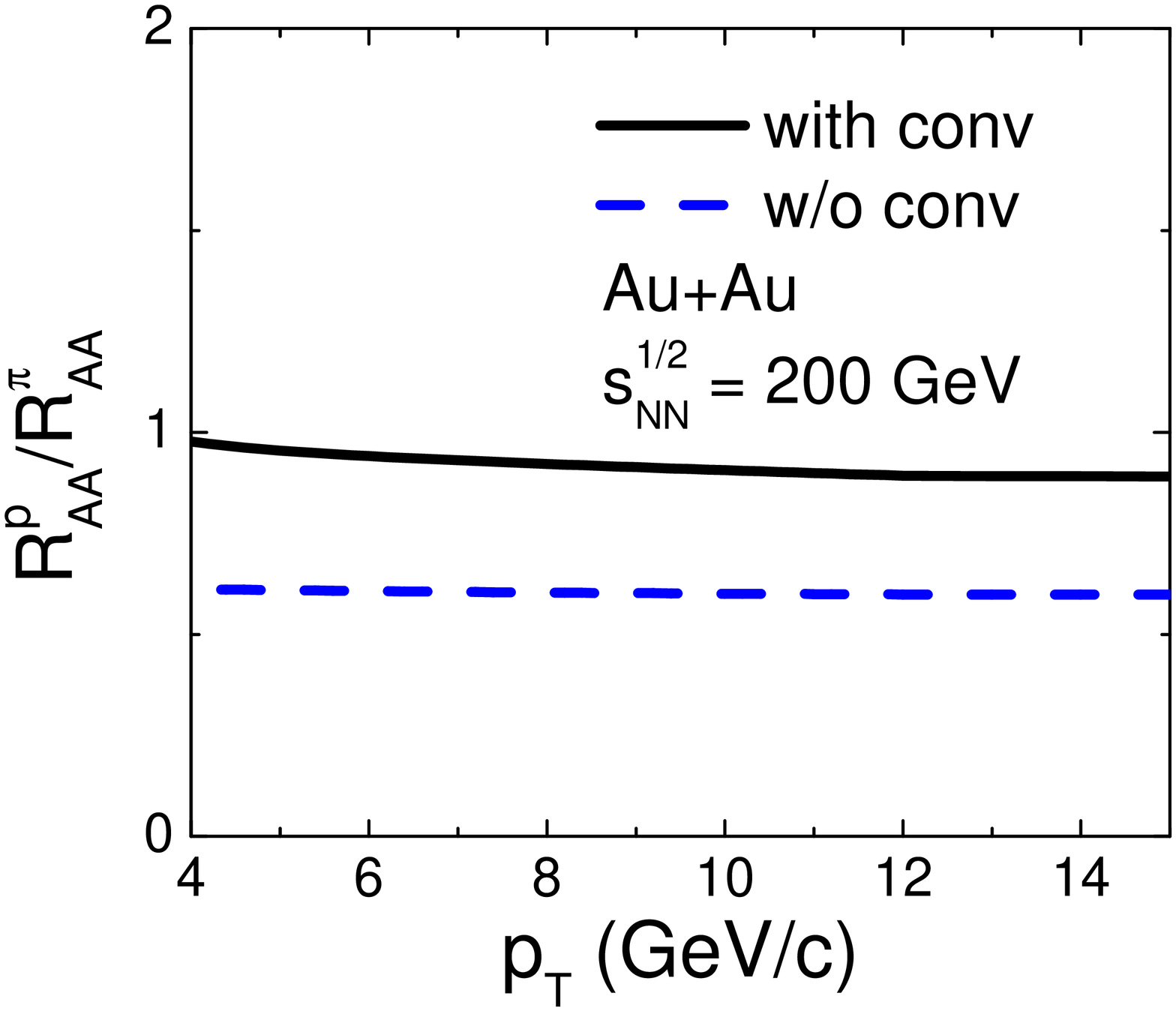}}
\caption{Left panel: The nuclear modification factor $R_{AA}$ for direct
photons with and without conversions switched on, calculated in the model
introduced in \cite{weiliu} (preliminary PHENIX data from \cite{Adler:2005ig}).
Right panel: The ratio of nuclear modification
factors for protons and pions is approaching one if conversions are allowed.}
\label{fig:1}
\end{figure}

Conversions between leading quark and gluon jet particles can potentially
answer the question why there are no signs of the additional color factor 
9/4 in the quenching of gluons. Hadrons that favor fragmentation from 
gluons more than others should exhibit smaller modification factors 
$R_{AA}$. The
data is pointing in the exactly opposite direction: protons are less 
suppressed than pions, even for momenta of 10 GeV/$c$ or more \cite{Xu:2009xy}.
It has been demonstrated that conversions can effectively
blur the distinction between well-defined quark or gluon jets.
The right hand side of Fig.\ \ref{fig:1} shows the increase
in proton $R_{AA}$ to the value of the pion $R_{AA}$ with conversions
computed in \cite{weiliu}. 
However, while conversions increase relative proton suppression to the
level of pions it can not explain the data from STAR which shows even 
less suppression for protons \cite{Xu:2009xy}. More work needs to go
into understand baryons in heavy ion collisions. It could be speculated 
that soft physics like quark recombination \cite{Fries:2003vb} is still
contributing to baryon production at those large momenta.

\begin{figure}[t]
\centerline{
\includegraphics[width=7.0cm]{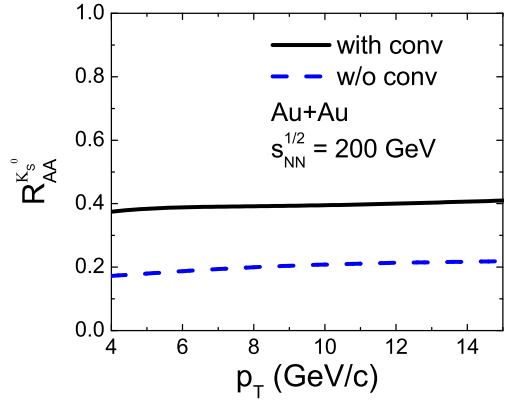}}
%\hspace{-0.2cm}
%\includegraphics[width=3.0in,height=3.0in,angle=-90]{photon_raa.eps}}
\caption{$R_{AA}$ for neutral kaons with and without conversion processes
allowed. The strangeness in the jet sample is driven towards equilibrium
by coupling it chemically to the quark gluon plasma.}
\label{fig:2}
\end{figure}

Historically, the first application of conversions of leading jet partons
was carried out for photon and dileptons. Annihilation and 
Compton processes of jets with partons from the medium can lead to hard 
photon emission 
\cite{fries1,FMS:05,SGF:02,GAFS:04}. This additional photons source is
now routinely integrated in state-of-the-art calculations of photon 
spectra \cite{SGF:02,simon,Turbide:2007mi}. Fig.\ \ref{fig:1}
shows the nuclear modification factor for direct photons from \cite{weiliu}.
Recently we also predicted that strange hadrons should be enhanced in
heavy ion collisions at RHIC energies \cite{weiliu}. 
Strange quark jets are rare at RHIC, while strange quarks are chemically 
equilibrated in the quark gluon plasma. Interactions between jets and the 
medium should therefore drive the jet ensemble toward chemical equilibration. 
through pair creation and kick out reactions of existing strange quarks 
in the medium.
This will translate into an increased $R_{AA}$ of kaons as 
shown in Fig.\ \ref{fig:2}.

We have checked that the same mechanism leads to a negligible enhancement
of heavy charm and bottom quarks at high momentum both at RHIC and LHC
energies \cite{Liu:2008bw}. The main reasons are the large thresholds 
for pair creation and the small amount of primordial heavy quarks in 
the medium which makes kick out reactions ineffective. 
%This assumes 
%that there is no chemical equilibration of heavy quarks even at LHC energies.

\begin{figure}[t]
\centerline{
\includegraphics[width=6.0cm,angle=-90]{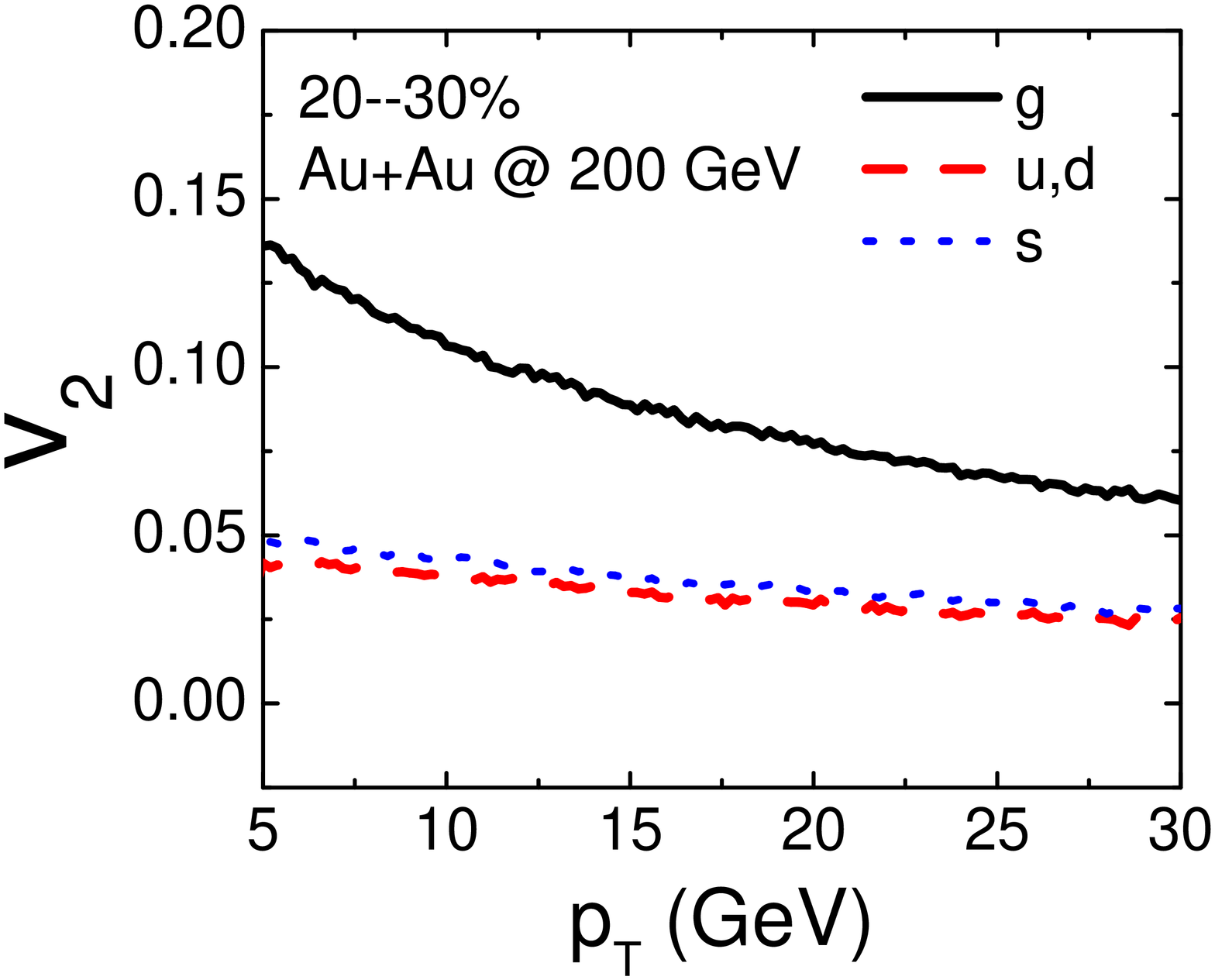}
%\hspace{-0.2cm}
\includegraphics[width=6.0cm,angle=-90]{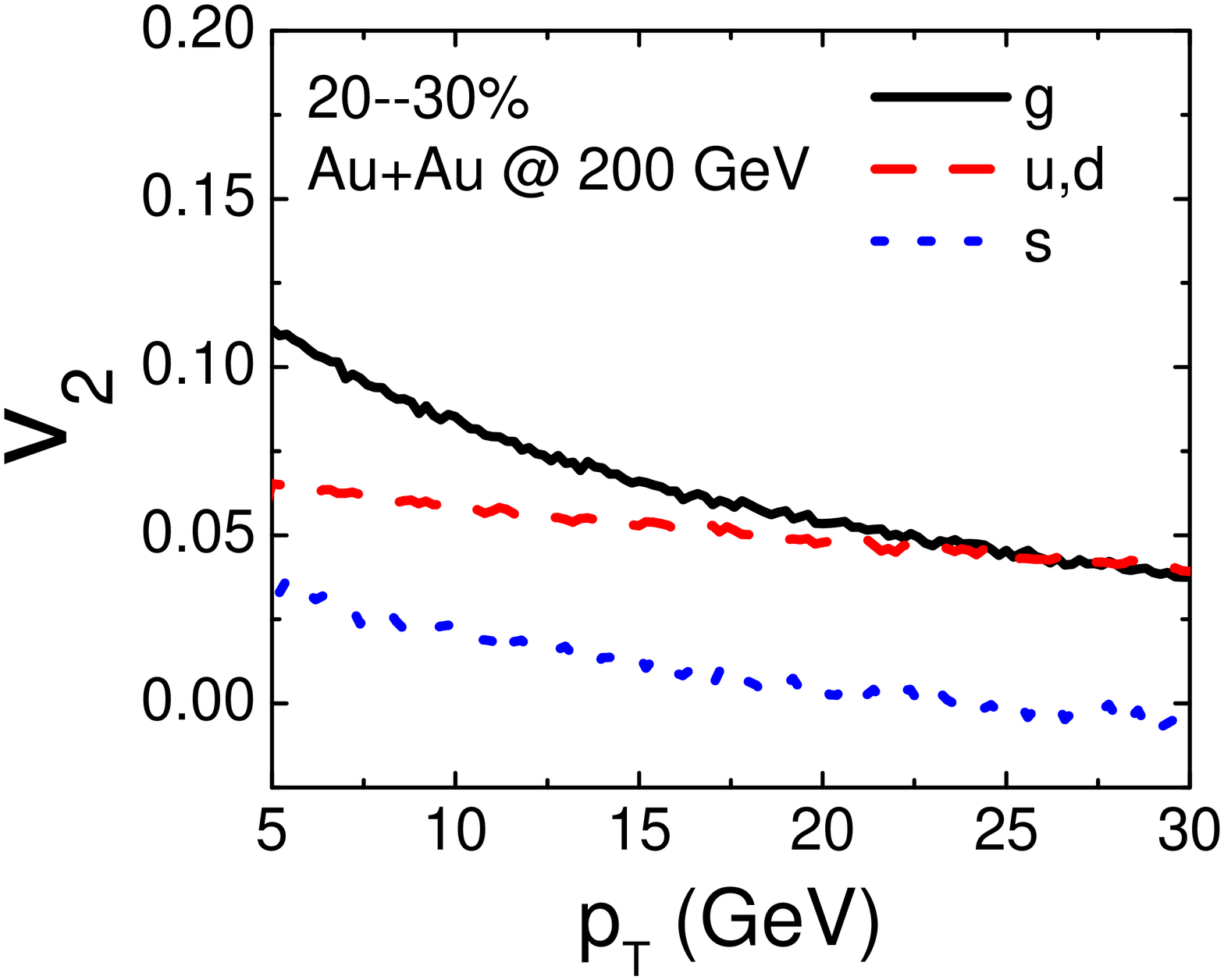}}
\caption{Left panel: The azimuthal asymmetry $v_2$ for light quarks, strange
quarks and gluons without conversions. Right panel: the same with conversions.
The $v_2$ for light quarks and gluons is not similar, while strange quarks
exhibit a suppression.}
\label{fig:3}
\end{figure}

\begin{figure}[t]
\centerline{
\includegraphics[width=6.5cm,angle=-90]{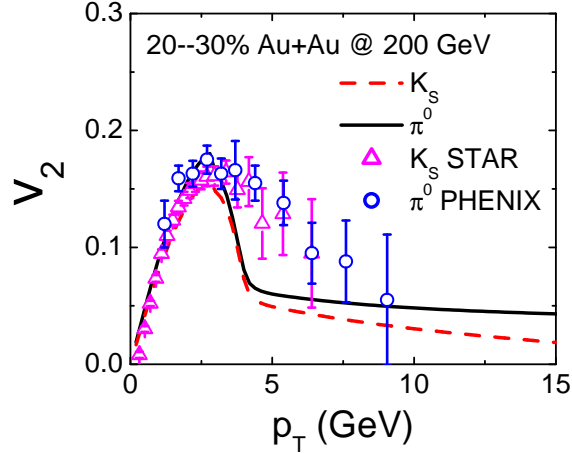}}
%\hspace{-0.2cm}
%\includegraphics[width=4.0cm,angle=-90]{quark_v2.eps}}
\caption{The resulting azimuthal asymmetry $v_2$ for kaons which is
expected to be suppressed compared to that of pions. Data from
\cite{david,abelev}.}
\label{fig:4}
\end{figure}

A more recent development have been considerations of the azimuthal 
asymmetry coefficient $v_2$ for particles emerging from jet conversions.
It was first pointed out for photons that there are more conversions if
a parton has to travel through thicker material, thus rendering the 
elliptic flow coefficient $v_2$ negative for the conversion product
\cite{Turbide:2005bz}. 
After adding photons from other sources the resulting $v_2$ is positive 
but numerically very small. First preliminary data from PHENIX on photon 
$v_2$ at large momentum is still inconclusive 
\cite{Turbide:2005bz,Chatterjee:2005de,weiliu}. 
We advocate similar efforts to measure the $v_2$ of strange hadrons at 
high $P_T$. Just like photons, strange quarks from jet conversions should
have negative $v_2$. Adding all sources we predict a suppression of the
$v_2$ for strange quarks compared to gluons and light quarks and hence a
suppression of the $v_2$ of kaons compared to pions \cite{Liu:2008kj}, 
see Fig.\ \ref{fig:3} and \ref{fig:4}.

%%%%%%%%%%%%%%%%%%%%%%%%%%%%%%%%%%%%%%%%%%%%%%%%
%% BACKMATTER
%%%%%%%%%%%%%%%%%%%%%%%%%%%%%%%%%%%%%%%%%%%%%%%%

%\begin{theacknowledgments}
  This work was supported by RIKEN/BNL Research Center and DOE grant
  DE-AC02-98CH10886.
%\end{theacknowledgments}

\end{document}